\documentclass[letterpaper, 10 pt, conference]{ieeeconf}  

\IEEEoverridecommandlockouts                              
\usepackage[utf8]{inputenc}
\usepackage[T1]{fontenc}
\usepackage{graphicx}
\usepackage{xcolor}

\usepackage[utf8]{inputenc} 
\usepackage[T1]{fontenc}    
\usepackage{hyperref}
\usepackage{url}
\usepackage{booktabs}       
\usepackage{array}
\usepackage{amsfonts}       
\usepackage{nicefrac}       
\usepackage{microtype}      
\usepackage{subfigure}
\usepackage{graphicx}
\usepackage{textcomp}
\usepackage{amssymb}
\usepackage{amsmath}
\usepackage{algorithm}
\usepackage{algorithmicx}
\usepackage{algpseudocode}
\usepackage{multirow}
\usepackage{amsmath}
\usepackage{amssymb}
\usepackage[symbol]{footmisc}
\usepackage{color}

\title{\LARGE \bf
FRSign: A Large-Scale Traffic Light Dataset for Autonomous Trains
}

\author{Jeanine Harb$^{1}$, Nicolas Rébéna$^{1}$, Raphaël Chosidow$^{2}$, Grégoire Roblin$^{3}$, Roman Potarusov$^{3}$ and Hatem Hajri$^{1}$
\thanks{$^{1}$J. Harb, N. Rébéna and H. Hajri are with Institut de Recherche Technologique SystemX, 91120 Palaiseau, France. Email:
        {\tt\small first name.second name@irt-systemx.fr}}%
\thanks{$^{2}$R. Chosidow is with Société Nationale des Chemins de Fer, 75014 Paris, France. Email:
        {\tt\small first name.second name@sncf.fr}}%
\thanks{$^{3}$G. Roblin and R. Potarusov are with Alstom, 93400 Saint-Ouen, France. Email:
        {\tt\small first name.second name@alstomgroup.com}}%
}

\begin{document}

\maketitle
\thispagestyle{empty}
\pagestyle{empty}

\begin{abstract}
In the realm of autonomous transportation, there have been many initiatives for open-sourcing self-driving cars datasets, but much less for alternative methods of transportation such as trains. 

In this paper, we aim to bridge the gap by introducing FRSign, a large-scale and accurate dataset for vision-based railway traffic light detection and recognition. 

Our recordings were made on selected running trains in France and benefited from carefully hand-labeled annotations. 

An illustrative dataset which corresponds to ten percent of the acquired data to date is published in open source with the paper. It contains more than 100,000 images illustrating six types of French railway traffic lights and their possible color combinations, together with the relevant information regarding their acquisition such as date, time, sensor parameters, and bounding boxes. This dataset is  published in open-source at the address \url{https://frsign.irt-systemx.fr}.

We compare, analyze various properties of the dataset and provide metrics to express its variability. We also discuss specific challenges and particularities related to autonomous trains in comparison to autonomous cars.
\end{abstract}

 
\section{Introduction}

Intelligent transport systems (ITS) present nowadays a major interest to thousands of companies all over the world. These systems encompass a wide range of technologies, from detection and recognition to telecommunications, traffic management and security, to mention a few. ITS are now at different stages of development with the aim to achieve a better organization of daily transport, provide better quality of service, add more safety and comfort for users, and improve coordination for traffic management. They can also save more time, cost or energy, and thus contribute to the development and design of tomorrow's digital city. 

Autonomous cars are certainly the means of transportation that have so far deserved the most resources, development, research and analysis. Developing autonomous cars is certainly the most challenging given the complexity of the road environment, particularly when many vehicles have to interact between each others. But cars are far from being the only vehicle of interest in the quest towards automation. Other modes of transportation have recently joined the race such as trains, ships and aircrafts. In this paper, our main concern will be in mainline trains. While automation of trains is less complex than that of cars and has been solved in closed environment already (e.g. automatic metros), the challenges remain significant in open environments. A first difficulty is to build an obstacle detection system with efficient sensors that is able to detect obstacles at long distance and which remains robust against external perturbations such as branches falling on the tracks, rockslides, animals and human intrusion. A second challenge is to design an automatic traffic light system which is able to recognize and properly interpret lights of existing railway signalization systems. All these technologies must respect the real-time constraint which is much more challenging in the field of autonomous trains than autonomous vehicles given trains speeds. Safety is another important component that should be considered seriously in the development of autonomous trains. Indeed, restrictions are much more severe in the world of autonomous trains than in the world of autonomous vehicles \cite{inproceedings}.

Even though trains are behind time in terms of vehicular automation, there is a crude sense of excitement when learning about the initiatives being spearheaded at national and international levels. For instance, France’s national railway company, Société Nationale des Chemins de Fer (SNCF), expects to see semi-autonomous trains running on the French rail network by 2020, and fully automatic trains within five years. In order to achieve full autonomy, trains will be equipped with advanced (and/or intelligent) cameras, high-precision sensors, and state-of-the-art positioning systems. These sensors are required for monitoring the railway and the train's environment. Therefore, they record a massive amount of data which, when acquired, play a critical role. Indeed, these data could one day enable computers to mimic a train driver's observational capabilities, through the training of machine learning models. On the other hand, having data available in public allows researchers and developers to work in a competitive environment, which can only accelerate technological development. 

Open sourcing data has proven effective in the development of autonomous vehicles \cite{journals/corr/JanaiGBG17}. For example, KITTI \cite{Geiger2012CVPR} can be considered as the first destination for autonomous driving researchers. It contains annotated data from scene flow, sensors and 3D object localization. The benchmarks on KITTI are a battleground for researchers in pursuit of the sleekest algorithms. BDD100K \cite{yu18} is probably the hugest publicly available self-driving dataset. It includes object detection, lane detection, drivable area and semantic segmentation sub-datasets. COCO \cite{lin2014microsoft} is a large-scale object detection, segmentation, and captioning dataset designed for object recognition. nuScenes \cite{nuscenes2019} is a more recent image-based benchmark dataset for object detection and tracking. These datasets, as  well as others, have been the driving force behind various developments in autonomous vehicles. 

\begin{figure}[!htp]
\centering
     \includegraphics[width=0.44\textwidth]{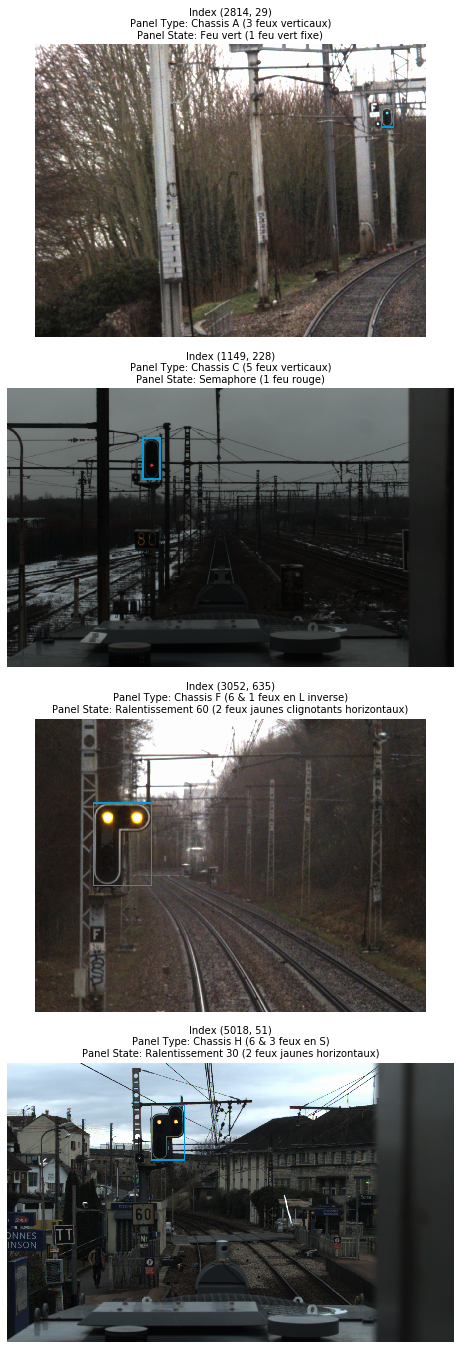}
      \caption{Images from FRSign Dataset}\label{fig:sample_images}
\end{figure}

In the field of autonomous trains, open sourcing data initiatives are still very limited. Recently, \cite{DBLP:conf/cvpr/ZendelMZSAB19} introduced RailSem19 as the first public dataset containing annotated images for semantic scene understanding in the rail environment. In the present paper, we introduce FRSign, the first dataset containing images for railway traffic lights. In addition to having benefited from manual annotation, our dataset is accompanied with several acquisition information which highlight its variability. Some samples of labeled railway traffic lights from FRSign are shown in Figure \ref{fig:sample_images}. Unlike the automotive sector, railway lights specifications are not unified, and each country designs and uses its own signals. Our dataset was acquired in France and contains multitudes of samples of the French railway traffic lights which will be described in the paper.  

The rest of the paper is organised as follows. Section \ref{sec3} discusses general challenges of autonomous trains in comparison with autonomous vehicles. Our dataset FRSign is introduced and described in Section \ref{sec4} together with several details regarding its acquisition and annotation. Some samples are also given and visualized. Section \ref{sec5} presents some statistics and metrics expressing the variability of the dataset. Finally in Section \ref{sec6}, we discuss possible interests of FRSign towards automation of the task of reading lateral signalling for train and present related specific challenges.





\section{Challenges of autonomous trains vs. autonomous cars}\label{sec3}

According to the International Association of Public Transport (UITP), there are four Grades of Automation (GoA) of trains \cite{uitp}:
\begin{enumerate}
    \item GoA1: The first grade is manual train operation, where the train driver controls starting and stopping, operation of doors and handling of emergencies or sudden diversions.
    \item GoA2: The second grad handles train starting and stopping operations.
    \item GoA3: The third grade is driverless train operation where there are no drivers, but an attendant is onboard in order to take control in case of  emergency.
    \item GoA4: Finally, the fourth grade corresponds to unattended train operation, which is true automation without any staff on board.
\end{enumerate}



Technical progress has made train control systems capable more than ever before of many complex tasks such as supervising, operating and controlling the entire operational process making autonomous trains possible.

Even if it might be thought that the presence of railways renders the challenge of autonomous trains easier than autonomous cars, multiple factors make the train a unique problem, which cannot be solved by a simple transposition of the solutions found for other types of vehicles. Here are some illustrative examples of challenges for trains:


\begin{figure*}[!htp]
\centering
     \includegraphics[width=0.9\textwidth]{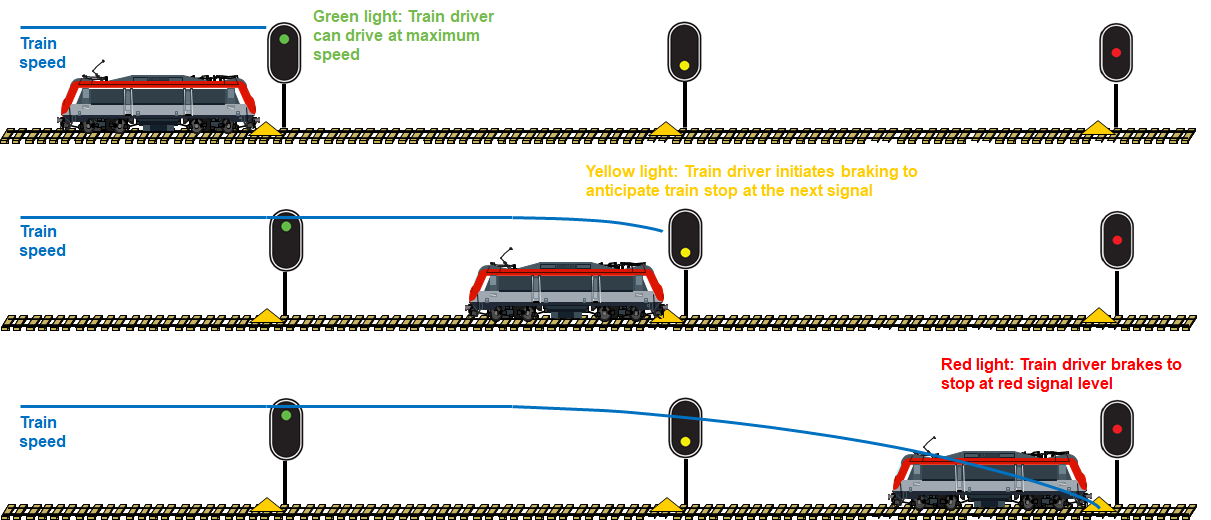}
      \caption{Rail Signalization Principle}\label{fig:principe_signalisation}
\end{figure*}

\begin{enumerate}
    \item One of the main difficulties lies in the complex interaction of the various subsystems when the train is running: monitoring the status of the railway track, the position of other trains and the physical integrity of the train, and determining the required space to brake safely. If any subsystem cannot judge the situation accurately, it will perform defensively by slowing or stopping, which hinders transport performance.

    \item The variability of use cases in the rail industry is much more important than for cars:
    \begin{enumerate}
        \item Railway signalling is diverse and features multiple types of signals/state combinations. For example, a railway panel of type ‘H’ (cf. Section \ref{sec:traffic_light_specs}) is configured to show up-to 18 different states, using eight light bulbs.
        \item There exists a plethora of train types with matching infrastructure. Thus, an autonomous system that applies to one type of trains may not necessarily generalize to another. While there is a will for unification, overhauling a country’s entire rail infrastructure is a daunting long-term project that comes at high cost.
    \end{enumerate}
    \item The stopping distance of a train is function of its body weight, cargo, and speed: a 10-car passenger train moving at 100 km/h requires about 500 meters to come to a halt. Therefore, it is imperative to detect obstacles and recognize the state of traffic lights well in advance before making a calculated stop.
    \item Due to the long stopping distances mentioned above, the railway signalling system is based on block sections protected by traffic lights. Contrary to road traffic where drivers must adapt the speed to anticipate changing state of the next visible traffic light (i.e. a driver adapts the braking effort to stop or not his/her vehicle at a traffic light depending on the speed and distance to the traffic light not to cross the red signal), rail signalization considers a succession of signals protecting successive block sections. The lengths of the block sections are defined so that any train respecting the signalization can stop at a stopping signal (red signal) safely. Typically, the train driver is informed by a yellow signal that the next signal is red. Thus, the train driver must brake and anticipate to stop at the next signal. Figure \ref{fig:principe_signalisation} illustrates this principle. Similarly, specific states of the traffic lights inform the driver in advance that a speed limit must be applied from the next signal.
    \item The railway signalization must strictly be respected no matter the weather conditions, even when visibility is very limited. An autonomous system must therefore demonstrate its capability to read signals in all conditions in a safe manner before the train crosses the signal.
    \item A special emphasis is given on passenger safety and system dependability. Without proof of the computer system’s reliability, security, and robustness, autonomous trains cannot be certified, let alone commercialized.
    \item Testing the system in situ requires access to railway infrastructure, which is not as trivial a matter compared to regular roads.
\end{enumerate}

These bottlenecks may be holding GoA4 trains back today, but there is no doubt that we are on the brink of a breakthrough that will revolutionize public transportation as we know it. Both freight and passenger trains will benefit from full autonomy: driverless operations increase system availability, network capacity and operational efficiency.

One thing is certain: it is but a matter of time before we reach industrial maturity in this area. Indeed, autonomous trains are expected to mark the beginning of the third “rail revolution”, after electric traction and high-speed rail. Moreover, they will cement AI’s reputation for being the biggest disruption of the modern era.

\section{Data acquisition and labeling}
\label{sec4}

The name of the dataset, FRSign, stands for "French Railway Signalling". The collection of the data released in FRSign took place on specific portions of railroads defined in Section \ref{sec:use_case}. We proceed by presenting the characteristics of French railway traffic lights in Section \ref{sec:traffic_light_specs}. We finally end by describing how we collected the data found in FRSign (Section \ref{sec:acquisition}), and how the images included in the dataset were annotated (Section \ref{sec:annotation}).

\subsection{Use Case Definition}
\label{sec:use_case}

\begin{figure}[!htp]
\centering
     \includegraphics[width=0.45\textwidth]{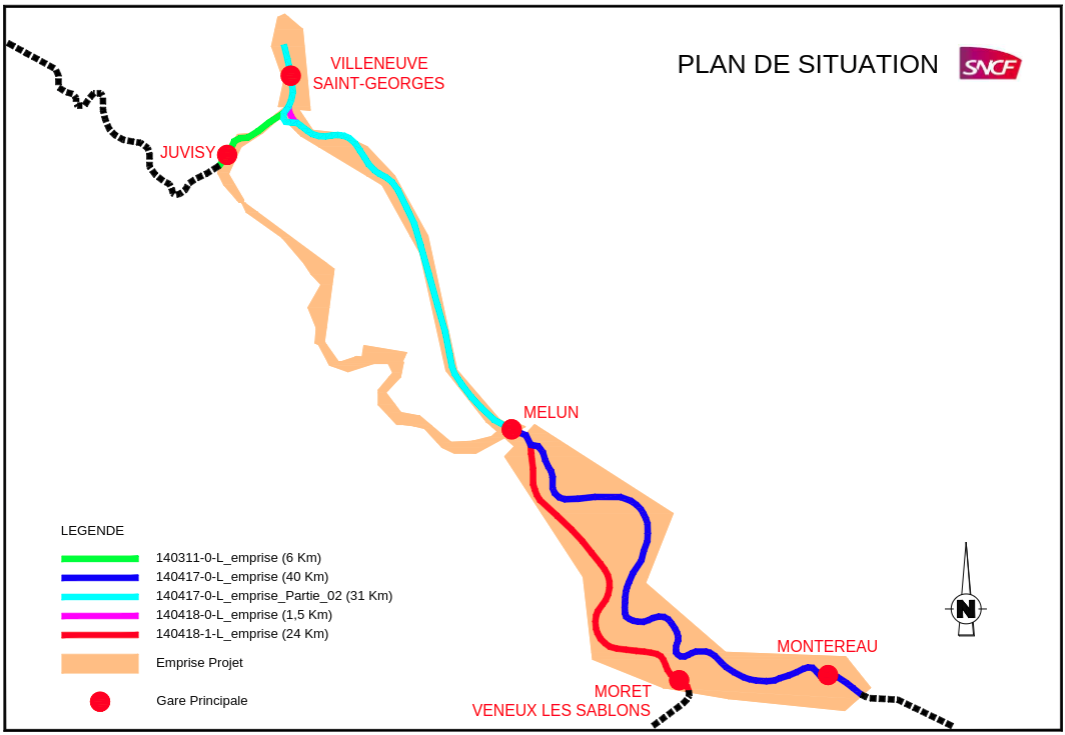}
      \caption{Use Case Definition}\label{fig:use_case}
\end{figure}

For the purposes of our work, our team had access to a train, portrayed in Figure \ref{fig:bb66000}, that circulates twice per month ever since September 2017 on specific railways located between the towns Villeneuve Saint-Georges and Montereau, in the region of Ile-de-France in France, highlighted in orange in Figure \ref{fig:use_case}. Most of the existing French railway panels can be found on these portions of railways, which is the reason why they were initially chosen for the use case.

\subsection{French Railway Traffic Lights}
\label{sec:traffic_light_specs}

\begin{figure}[!htp]
\centering
     \includegraphics[width=0.45\textwidth]{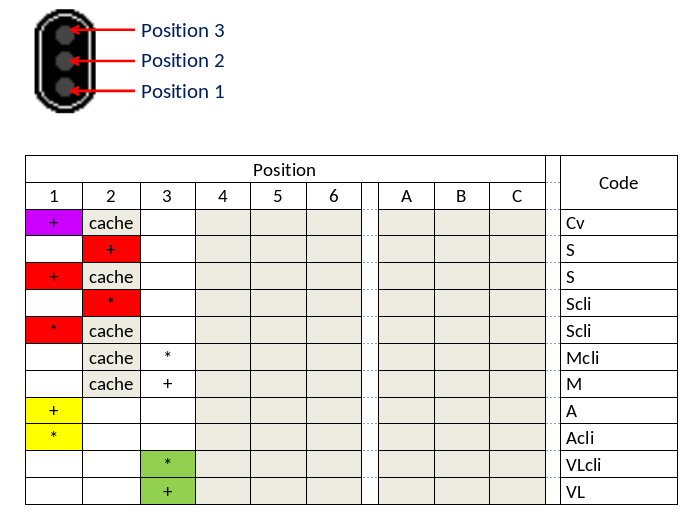}
      \caption{Panel A Combinations}\label{fig:panel_a_combinations}
\end{figure}

\begin{figure}[!htp]
\centering
     \includegraphics[width=0.45\textwidth]{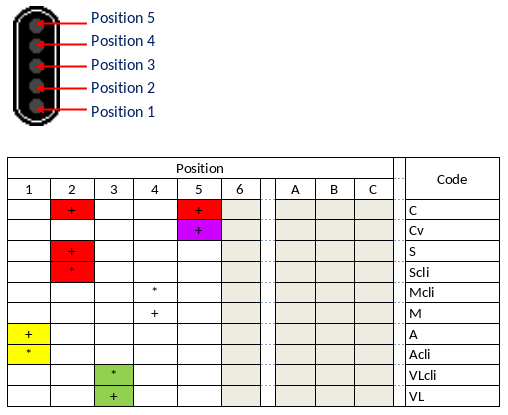}
      \caption{Panel C Combinations}\label{fig:panel_c_combinations}
\end{figure}

In FRSign, we selected a subset of French railway panels that were the most recorded during our tests. The list hereafter presents the six types of French railway panels included in the dataset and describes their configuration:
\begin{itemize}
    \item \textit{'Chassis A (3 feux verticaux)'}: panel named A featuring three vertical lights (cf. Figure \ref{fig:panel_a_combinations}).
    \item \textit{'Chassis C (5 feux verticaux)'}: panel named C featuring five vertical lights (cf. Figure \ref{fig:panel_c_combinations}).
    \item \textit{'Chassis F (6 \& 1 feux en L inverse)'}: panel named F featuring six vertical lights plus one light in an inverse L-shaped composition.
    \item \textit{'Chassis H (6 \& 3 feux en S)'}: panel named H featuring six vertical lights plus three lights in an S-shaped composition.
    \item \textit{'Chassis ID2 (2 feux horizontaux)'}: panel named ID2 featuring two horizontal lights.
    \item \textit{'Chassis ID3 (3 feux horizontaux)'}: panel named ID3 featuring three horizontal lights.
\end{itemize}

Table \ref{tab:panel_states} contains the possible states that can be found in the dataset, ordered from most restrictive to most permissive. Note that these descriptions apply only to panels A, C, F and H.

\begin{table*}[ht]
\centering
\begin{tabular}{|l|l|l|}
\hline
\textbf{State name in dataset}                                       & \textbf{Code} & \textbf{State description}                           \\ \hline
\textit{'Carre (2 feux rouges)'}                                     & C             & Full stop, two red lights                            \\ \hline
\textit{'Carre violet (1 feu violet)'}                               & Cv            & Maintenance stop, single purple light                \\ \hline
\textit{'Semaphore (1 feu rouge)'}                                   & S             & Stop, single red light                               \\ \hline
\textit{'Feu rouge clignotant (1 feu rouge clignotant)'}             & Scli         & Blinking stop, single blinking red light             \\ \hline
\textit{'Feu blanc (1 feu blanc)'}                                   & M             & Maneuver            \\ \hline
\textit{'Avertissement \& rappel 30'}                                &  RR+A         & Warning (single yellow light) and speed limit 30.    \\ \hline
\textit{'Feu jaune clignotant \& rappel 30'}                         &  RR+Acli      & Blinking yellow light and speed limit 30.            \\ \hline
\textit{'Rappel 30 (2 feux jaunes verticaux)'}                       & RR            & Speed limit 30, two vertical yellow lights.          \\ \hline
\textit{'Avertissement \& rappel 60'}                                &  RRcli+A      & Warning (single yellow light) and speed limit 60.    \\ \hline
\textit{'Feu jaune clignotant \& rappel 60'}                         &  RRcli+Acli  & Blinking yellow light and speed limit 60.            \\ \hline
\textit{'Rappel 60 (2 feux jaunes clignotants verticaux)'}           & RRcli        & Speed limit 60, two blinking vertical yellow lights. \\ \hline
\textit{'Avertissement (1 feu jaune)'}                               & A            & Warning, single yellow light.      \\ \hline
\textit{'Ralentissement 30 (2 feux jaunes horizontaux)'}             & R            & Slow down 30, two horizontal yellow lights.          \\ \hline
\textit{'Feu jaune clignotant \& ralentissement 60'}                 & Rcli+Acli    & Yellow light and slow down 60                        \\ \hline
\textit{'Feu jaune clignotant (1 feu jaune clignotant)'}             & Acli         & Single blinking yellow light                         \\ \hline
\textit{'Ralentissement 60 (2 feux jaunes clignotants horizontaux)'} & Rcli         & Slow down 60, two blinking horizontal yellow lights  \\ \hline
\textit{'Feu vert clignotant (1 feu vert clignotant)'}               & VLcli        & Slow down 160, single blinking green light           \\ \hline
\textit{'Feu vert (1 feu vert fixe)'}                                & VL           & Go, single green light                               \\ \hline
\end{tabular}
\caption{Railway Traffic Light States for Panels A, C, F and H}
\label{tab:panel_states}
\end{table*}


As for panels ID2 and ID3, their function is not to regulate railway traffic, but to indicate directions. Their states are explained in Table \ref{tab:id_states}:

\begin{table}[ht]
\centering
\begin{tabular}{|l|l|}
\hline
\textbf{State name in dataset}              & \textbf{State description}       \\ \hline
\textit{'Feu blanc (1 feu blanc)'}          & Direction 1, single white light             \\ \hline
\textit{'Feux blancs (2 feux blancs)'}      & Direction 2, two white lights             \\ \hline
\end{tabular}
\caption{Railway Traffic Light States for Panels ID2 and ID3}
\label{tab:id_states}
\end{table}

\begin{figure}[!htp]
\centering
     \includegraphics[width=0.45\textwidth]{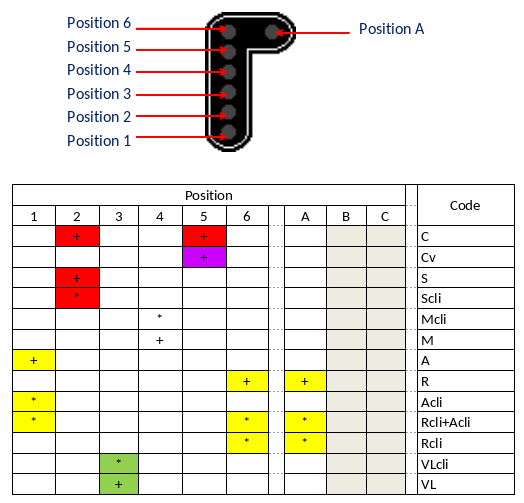}
      \caption{Panel F Combinations}\label{fig:panel_f_combinations}
\end{figure}

\begin{figure}[!htp]
\centering
     \includegraphics[width=0.45\textwidth]{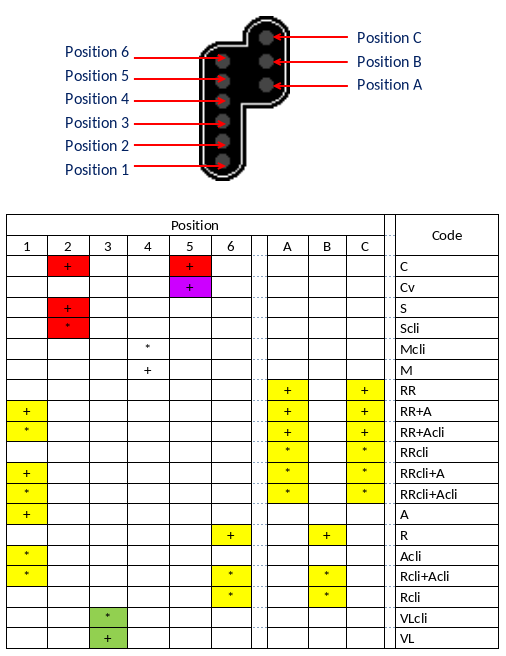}
      \caption{Panel H Combinations}\label{fig:panel_h_combinations}
\end{figure}

\begin{figure}[!htp]
\centering
     \includegraphics[width=0.45\textwidth]{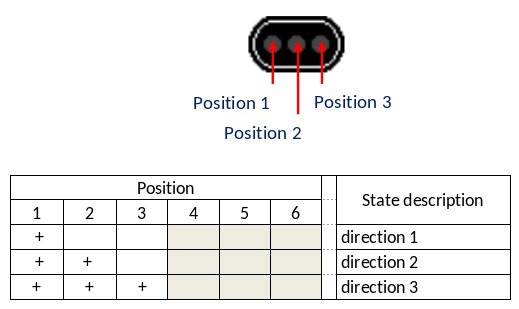}
      \caption{Panel ID Combinations}\label{fig:panel_id_combinations}
\end{figure}

All states do not apply to all panels. The possible light combinations for each panel are explained in figures \ref{fig:panel_a_combinations}, \ref{fig:panel_c_combinations}, \ref{fig:panel_f_combinations}, \ref{fig:panel_h_combinations}, and \ref{fig:panel_id_combinations}. In each of these figures, the symbol "+" corresponds to a fixed light, and the symbol "*" corresponds to a blinking light. "Cache" means that the light bulb is not being used or is hidden behind a blocker. Also, note that for each panel, we may have multiple possible color configurations: for instance, panel A may either have a violet, red, or yellow light bulb in "Position 1". 

\subsection{Data Acquisition}
\label{sec:acquisition}

\begin{figure}[!htp]
\centering
     \includegraphics[width=0.45\textwidth]{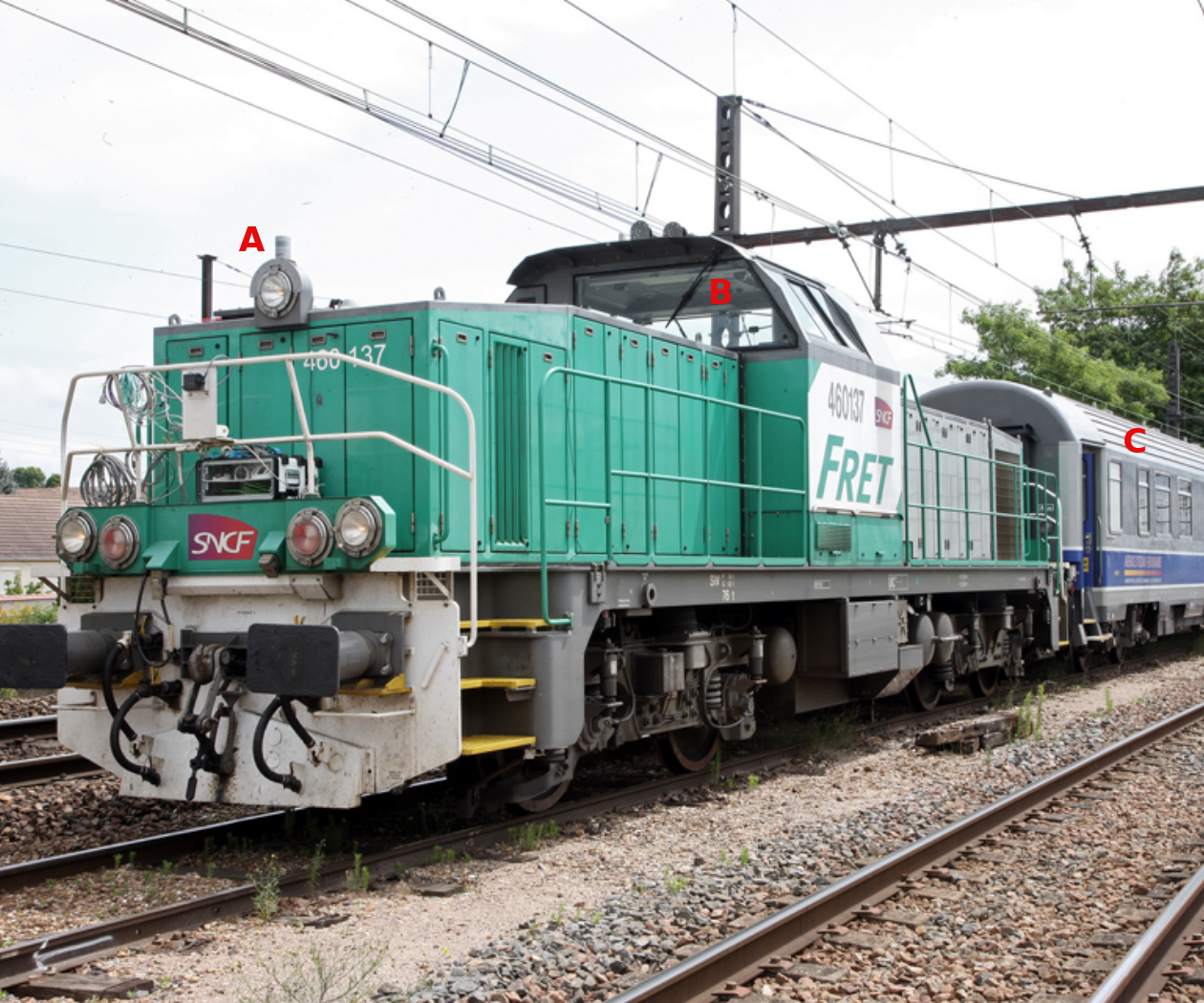}
      \caption{Train used to acquire the dataset}\label{fig:bb66000}
\end{figure}

The train used for our work is equipped with multiple sensors that detect and capture the surrounding environment. For instance, it is equipped with a LiDAR fixed on top of the train's engine (A in Figure \ref{fig:bb66000}). Most importantly, it is equipped with two cameras, located in the driver's cabin (B in Figure \ref{fig:bb66000}), that capture the railway in front of the train. Therefore, these cameras film the railway from the train driver's point of view. In the dataset, they are identified by unique identifiers (or \texttt{sensor\_id}): \texttt{camera\_1} and \texttt{camera\_2}. The specifications of the two cameras are summarized in Table \ref{tab:camera_specs}. The acquired images are then stored in an embarked lab in the test vehicle behind the locomotive (C in Figure \ref{fig:bb66000}).

\begin{table}[ht]
\centering
\begin{tabular}{|l|l|l|}
\hline
\textbf{Camera ID}     & camera\_1 & camera\_2 \\ \hline
\textbf{Megapixels}    & 3.2       & 2.3       \\ \hline
\textbf{Sensor Format} & 1/1.8"    & 1/1.2"    \\ \hline
\textbf{Pixel Size}    & 3.45 µm   & 5.86 µm   \\ \hline
\textbf{Resolution}    & 2048x1536 & 1920x1200 \\ \hline
\textbf{Maximum Frame Rate}    & 121 FPS   & 163 FPS   \\ \hline
\end{tabular}
\caption{Cameras Specifications}
\label{tab:camera_specs}
\end{table}

\subsection{Image Annotation}
\label{sec:annotation}

The recordings made on the train were subsequently annotated thanks to the software BeaverDam \cite{Shen:EECS-2016-193}, which we have adapted to our use. More precisely, we used two levels for labeling: classes refer to the railway panel's type, and subclasses refer to the panel's current state. Therefore, we were able to annotate sequences of images, with each sequence corresponding to an individual state for the observed panel.

\newpage
\section{Dataset content and structure}\label{sec5}

We define in this section the content and the structure of the released dataset, with an emphasis on the terminology used to differentiate sequences and images.

\subsection{Video Sequence vs. Images}

In FRSign, we oppose two notions:
\begin{itemize}
    \item Video sequences: sequences of consecutive images showing an annotated video portion of a railway panel, continuously featuring the same state.
    \item Images: individual images of railway panels.
\end{itemize}

\subsection{Bounding boxes}

Annotations are implemented in the form of objects called "bounding boxes". A bounding box defines the region of interest in the acquired images. In our dataset, it delimits the railway traffic light we would like to detect, and is defined using four coordinates: $x, y, w, h$. 

\begin{figure}[!htp]
\centering
     \includegraphics[width=0.4\textwidth]{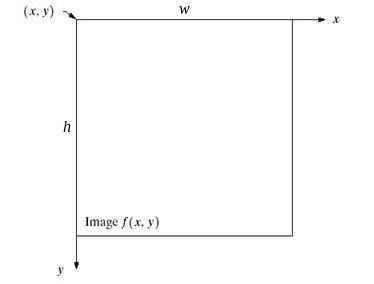}
      \caption{Bounding box implementation}\label{fig:boundingbox}
\end{figure}

As seen in Figure \ref{fig:boundingbox}, the tuple $(x, y)$ designates the upper left corner of the bounding box. The remaining coordinates, $w$ and $h$, define the width and the height of the rectangle drawn from the upper left corner.

\subsection{Content}

The dataset FRSign contains two types of files:
\begin{enumerate}
    \item An HDF5 file called "frsign\_v1.0.h5": this file lists all the images contained in the dataset with their corresponding metadata. The structure of the HDF5 file is explained hereafter.
    \item Image files: all the images have been saved in PNG format. Their descriptions can be found in the HDF5 file.
\end{enumerate}

The HDF5 file "frsign\_v1.0.h5" contains two dataframes named: 
\begin{enumerate}
    \item \texttt{/dataframe} : dataframe which contains the metadata related to all video sequences. 
    \item \texttt{/images} : dataframe with a double index that lists images according to their chronological order in a sequence as well as their corresponding bounding boxes. 
\end{enumerate}

Hereafter is the description of the columns found in  \texttt{/dataframe}:

\begin{itemize}
    \item \texttt{camera}: Identifier of the camera that was used.
    \item \texttt{CameraInfo\_bayerTileFormat}: Bayer tile format.
    \item \texttt{CameraInfo\_sensorResolution}: Sensor resolution (\textit{width} x \textit{height}).
    \item \texttt{context}: Context of the recording (train in our case).
    \item \texttt{datetime}: Date and timestamp of the recording.
    \item \texttt{fps}: Frame per second.
    \item \texttt{image\_format}: Image format (PNG8 in our case).
    \item \texttt{on\_track}: Boolean that indicates whether the railway panel is on the same track as the train.
    \item \texttt{optic}: Selected optic for the acquisition.
    \item \texttt{sensor\_id}: Unique sensor identifier.
    \item \texttt{sensor\_type}: Type of sensor that was used (camera in our case).
    \item \texttt{state}: State of the railway panel.
    \item \texttt{type}: Type of the railway panel.    \item \texttt{video\_name}: Name of the video file.
    \item \texttt{video}: Video filepath. 
\end{itemize}

The index of \texttt{/dataframe} creates a unique identifier for each video sequence found in the table.

Hereafter is the description of the columns found in   \texttt{/images}:

\begin{itemize}
    \item \texttt{fullpath} : Image filepath.
    \item \texttt{x} : Abscissa of the upper left corner of the bounding box.
    \item \texttt{y} : Ordinate of the upper left corner of the bounding box.
    \item \texttt{w} : Width of the bounding box. 
    \item \texttt{h} : Height of the bounding box. 
\end{itemize}

The double index of \texttt{/images} serves to, first, identify the video sequence that contains the image, and second, create a unique identifier for the image.

\section{Statistics}\label{sec6}

In this section, we present various statistics on the data found in the dataset FRSign.

To start, FRSign is composed of 393 video sequences, which make up 105,352 individual images. 

\begin{figure*}[!htp]
\centering
     \includegraphics[width=1.0\textwidth]{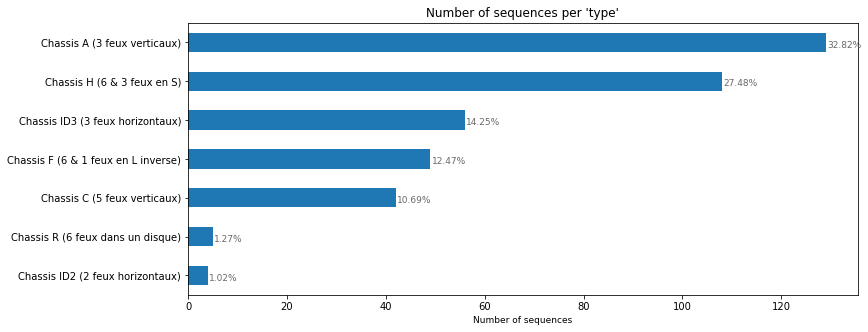}
      \caption{Number of video sequences per panel type}\label{fig:seq_per_type}
\end{figure*}

\begin{figure*}[!htp]
\centering
     \includegraphics[width=1.0\textwidth]{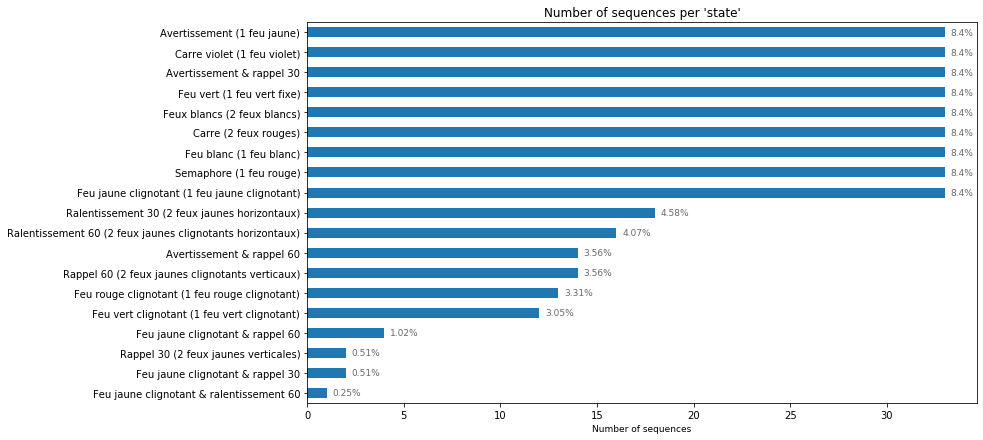}
      \caption{Number of video sequences per panel state}\label{fig:seq_per_state}
\end{figure*}

\begin{figure*}[!htp]
\centering
     \includegraphics[width=1.0\textwidth]{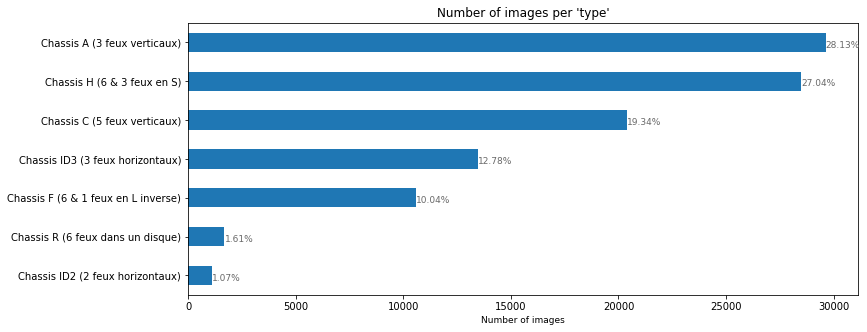}
      \caption{Number of images per panel type}\label{fig:img_per_type}
\end{figure*}

\begin{figure*}[!htp]
\centering
     \includegraphics[width=1.0\textwidth]{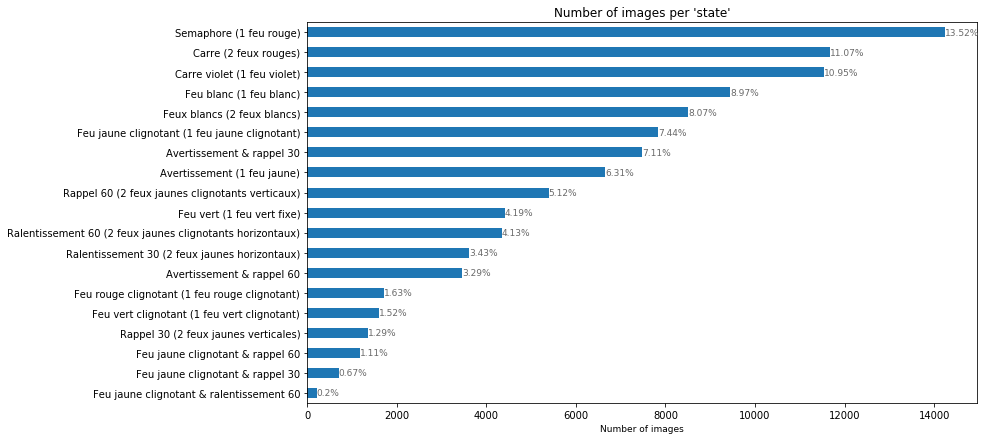}
      \caption{Number of images per panel state}\label{fig:img_per_state}
\end{figure*}

The horizontal bar charts found in Figures \ref{fig:seq_per_type}, \ref{fig:seq_per_state}, \ref{fig:img_per_type}, \ref{fig:img_per_state}, present the distribution of video sequences or individual images per panel type and state.

The displayed figures highlight the variability of the encountered railway traffic lights on the selected use case, respectively according to panel type and state.

\section{Conclusions and future work}\label{sec7}


In this paper, we introduced FRSign, a unique dataset containing more than 100,000 images of French railway signalling panels. We defined the terminology and the specifications used for French railway infrastructure such as panel types and their combinations.

Acquiring this data was not trivial. We faced multiple challenges on several levels:
\begin{itemize}
    \item Having access: having the necessary authorizations to access the railroads defined in our use case required a lot of preparation and organization.
    \item Configuring the hardware: using the correct camera parameters is of high importance for acquiring images of good. Indeed, we tested multiple optics with regards to the distance to the panels in order to visualize them well. Also, it was critical to find the best spot to fix cameras: we wanted to be on the same level as the train driver's eyes in order to have the same angle of view, and we wanted to avoid being near the engine, as it would make cameras vibrate.
    \item Understanding the panels: most captured panels use LED lights which required rectified voltage with a frequency of 100 Hz. If the camera's shutter speed is less than the LED's frequency, the acquired video sequences would falsely show a blinking light. Therefore, we understood that we needed to set the exposure time more than or equal to the frequency.
\end{itemize}

This dataset can be used to learn automatic classification of French railway signalling panels, either by type or by state. It could also be used to detect panels in the images by comparing the prediction to the supplied bounding box.

This work is a first attempt to open-sourcing French railway signalling panels. It can be completed and improved in various ways:
\begin{itemize}
    \item Acquire images under multiple different weather conditions: rain, snow, fog...
    \item Acquire images of all existing types of French railway signalling panels, as well as all their possible states, and reach a balanced distribution.
    \item Acquire images on railroads across the entire country, to cover all settings and environments.
\end{itemize}

We hope to address these challenges in future works.







\textbf{Acknowledgments.} This work was done at IRT SystemX as part of the project TAS (Safe Autonomous Land Tranport), in partnership with SNCF and Alstom. We would like to thank our colleagues: Loïc Cantat, Pierre d'Aviau De Ternay and Jack Fazakerley for their help in designing and setting up the website.

\bibliographystyle{IEEEtran}
\bibliography{reference}

\end{document}